\documentstyle[preprint,aps,prc,floats,epsfig]{revtex}
\tighten
\let\jnfont=\rm
\def\NPB#1,{{\jnfont Nucl.\ Phys.\ }{\bf B#1},}
\def\PLB#1,{{\jnfont Phys.\ Lett.\ B }{\bf #1},}
\def\PRD#1,{{\jnfont Phys.\ Rev.\ D }{\bf #1},}
\def\PRL#1,{{\jnfont Phys.\ Rev.\ Lett.\ }{\bf #1},}
\def\ZPC#1,{{\jnfont Z.~Phys.\ C }{\bf #1},}
\def\gev{\rm GeV}
\def\etmiss{{\overlay{/}{E}}_T}
\def\ptmiss{{\overlay{/}{p}}_T}
\def\lsim{\mathrel{\mathpalette\oversim<}}
\def\gsim{\mathrel{\mathpalette\oversim>}}
\def\oversim#1#2{\lower0.5ex\vbox{\baselineskip0pt\lineskip0pt
  \lineskiplimit0pt\everycr{}\tabskip0pt
  \halign{$\mathsurround0pt #1\hfil##\hfil$\crcr #2\crcr\sim\crcr}}}
\begin{document}
\draft
\preprint{} 

\preprint{
\vbox{\hbox{\bf MADPH--03--1358}
      \hbox{\bf TU-701}
      \hbox{\bf hep-ph/0312129}
}}
 
\title{ The FCNC top-squark decay as a probe of squark mixing}

\author{\ \\[2mm] Tao Han$^{1,2}$, Ken-ichi Hikasa$^3$, 
Jin Min Yang$^{2,3}$, Xinmin Zhang$^4$ } 

\address{ \ \\[2mm]
 $^1$ {\it Department of Physics, University of Wisconsin, 
               Madison, WI 53706, USA}\\
 $^2$ {\it Institute of Theoretical Physics, Academia Sinica, 
           Beijing 100080, China}\\
 $^3$ {\it Department of Physics, Tohoku University,
               Aoba-ku, Sendai 980-8578, Japan}\\
 $^4$ {\it Institute of High Energy Physics, Academia Sinica, 
           Beijing 100039, China }}

\maketitle

\begin{abstract}
In supersymmetry (SUSY) the flavor mixing between top-squark (stop) 
and charm-squark (scharm) 
induces the flavor-changing neutral-current (FCNC) stop decay 
$\tilde t_1 \to c \tilde \chi^0_1$.
Searching for this decay serves as a probe of soft SUSY breaking
parameters. Focusing on the stop pair production followed by the FCNC decay 
of one stop and the charge-current decay of the other stop, we investigate 
the potential of detecting this FCNC stop decay at the Fermilab Tevatron, 
the CERN Large Hadron Collider (LHC) and the next-generation $e^+e^-$ linear 
collider (LC). We find that this decay may not be accessible at the Tevatron, 
but could be observable at  the LHC and the LC  with  high sensitivity
for some part of parameter space.  
\end{abstract}


\section{ Introduction} 

Flavor-changing neutral-current  (FCNC) interactions are strongly 
suppressed in the Standard Model (SM) by the GIM mechanism, which
is consistent with the current experimental observation. In theories
beyond the SM the FCNC interactions are not generally suppressed, 
and thus are subject to stringent constraints from experiments \cite{fcp}. 
On the other hand,  the study of FCNC interactions, especially related 
to the top quark  \cite{hpz}, will play an important 
role in testing the SM and probing new physics.

Weak scale supersymmetry (SUSY) as a leading candidate for
new physics beyond the SM provides no further understanding
about the origin of flavor. In fact, it extends the mystery of flavor
by necessarily adding three families of squarks and sleptons.
Without additional assumptions for flavor structure of the soft SUSY
breaking, supersymmetric theories often encounter phenomenological
difficulties, known as the SUSY flavor problem \cite{susyf}.
This in turn implies that
there may exist rich FCNC phenomenology. Some highly suppressed 
FCNC processes in the SM may be enhanced in 
supersymmetric models to a level accessible in the 
future experiments, such as  $t\to cV$ ($V=\gamma, Z,g$) 
and $t\to ch$ \cite{tcvh_sm,tcv_mssm,tch_mssm}.
On the other hand, sfermions may have large flavor mixings 
via the soft SUSY breaking terms.
Even if the flavor-diagonality is assumed for sfermions at the 
grand unification scale,
the flavor mixings at weak scale are naturally 
generated through renormalization group equations \cite{duncan}. 
Therefore, hunting for the exotic FCNC processes predicted 
by SUSY would be one of the important aspects in SUSY 
searches at the upcoming colliders.

There have been intensive studies for the FCNC phenomenology 
in the slepton sector \cite{hall}. In the squark sector, some
interesting FCNC phenomena may arise from the mixing between 
the stop and scharm. 
On the experimental side, we note that despite of the strong 
constraints on the mixing between first and second generation squarks 
from $K^0$--$\bar K^0$ mixing, the mixing between stop and scharm is 
subject to less low-energy constraints and could be maximal \cite{review}
although a recent analysis \cite{Endo} of electric dipole moment of 
mercury atom indicated a nontrivial constraint on 
the $\tilde t_L-\tilde c_L$ mixing.
Such a large mixing would 
reveal itself in some processes or subject to some constraints in future 
collider experiments. On the theoretical side, in the Minimal Supersymmetric
Standard Model (MSSM) the stop-scharm 
mixing is likely to be large even if there is no mixing at tree level, 
as first realized in \cite{hikasa}. 
The stop-scharm mixing induces the FCNC stop decay 
$\tilde t_1 \to c \tilde \chi^0_1$, where $\tilde t_1$ is the 
lighter one of the stop mass eigenstates and 
$\tilde \chi^0_1$ is the lightest neutralino assumed 
to be the lightest supersymmetric particle (LSP).
Early searches for this channel at the Tevatron experiments 
have set the bounds $ m_{\tilde t_1} \gsim 120$ GeV for 
$m_{\tilde \chi^0_1}\sim 40$ GeV \cite{FCNC}. However, 
if kinematically accessible, the tree-level charged-current (CC) decay mode
$\tilde t_1 \to b \tilde \chi^+_1$, where $\tilde \chi^+_1$ is
the lighter chargino, will be most likely dominant although it may be
via a three-body decay with  $\tilde \chi^{+*}_1 \to \ell^+ \tilde\nu$
\cite{stop-3-body}.
Experimental searches for this mode have also been performed 
at the Tevatron  \cite{FCCC}, and the current bounds are
$ m_{\tilde t_1} \gsim 135$ GeV for  $m_{\tilde \nu}\sim 80$ GeV.
In their analyses, however, the decay branching fraction of $\tilde t_1$
has been simply assumed to be $100\%$ for each channel under 
their consideration. There have also been studies \cite{hosch} 
on the possibility of finding the FCNC stop decay from top 
quark pair production followed by the decay 
$t\to \tilde t_1 \chi^0_1\to  c \tilde \chi^0_1  \tilde \chi^0_1$
of one top and the SM decay of the other top.  
It is shown that such a decay mode, if realized in the $t\bar t$ 
pair events with a substantial branching fraction, could be 
observable in some part of the SUSY parameter space 
at Run 2 of the Tevatron collider \cite{cdf-tt}. 

In this article, we focus on the direct stop pair production followed by the 
FCNC decay of one stop ($\tilde t_1 \to c \tilde \chi^0_1$) and the 
charge-current decay of the other stop ($\tilde t_1 \to b \tilde \chi^+_1$).
We allow arbitrary branching fractions for these two channels.
By simulating both the signal and the SM backgrounds, we examine
to what levels the branching ratio and the stop-scharm mixing parameter 
can be probed at the Tevatron collider, the CERN Large Hadron Collider (LHC) 
and the 500 GeV next-generation $e^+e^-$ linear collider (LC).

\section{ Stop-scharm Mixing and FCNC Stop Decay}

Following Ref.~\cite{hikasa}, we start in the framework of the MSSM 
and assume that the tree level interactions are flavor
diagonal for stops and scharms. The flavor mixing between stops and 
scharms is then induced via loops. The dominant effects are from the
logarithmic divergences caused by soft breaking terms. Such divergences
must be subtracted using a soft counter-term at the SUSY breaking scale, 
such as the Plank scale $M_p$ in supergravity (SUGRA) models. Thus a large 
logarithmic  factor $(1/16\pi^2)\ln (M_p^2/m_W^2)\approx 0.5$ remains 
after renormalization\footnote{For some mechanisms of 
SUSY breaking other than gravity mediation,
such as gauge mediation, the SUSY breaking scale can be much lower than 
the Plank scale and thus this factor may be smaller.}. 
In the approximation of neglecting 
the charm quark mass, $\tilde c_R$ does not mix with stops. The mixing
of $\tilde c_L$ with stops result in the physical states given approximately 
by 
\begin{eqnarray} \label{mix}
\left (\begin{array}{l} 
       \tilde t_1\\ \tilde t_2\\ \tilde c_L
       \end{array} \right )_{\rm phys}=\left (\begin{array}{lll} 
                                               1 & 0 & \epsilon \\
                                               0 & 1 & \epsilon' \\
                                              -\epsilon & -\epsilon' &1 
                                                \end{array} \right )
               \left (\begin{array}{l} 
\tilde t_1\\ \tilde t_2\\ \tilde c_L                    \end{array} \right ),
\end{eqnarray}
where 
\begin{eqnarray}
\label{eps}
\epsilon = \frac{\Delta_L \cos\theta_t+\Delta_R \sin\theta_t}
 {m^2_{\tilde t_1}-m^2_{\tilde c_L}},\qquad
\epsilon' = \frac{\Delta_R \cos\theta_t-\Delta_L \sin\theta_t}
 {m^2_{\tilde t_2}-m^2_{\tilde c_L}},
\end{eqnarray}
with $\Delta_{L,R}$ given by 
\begin{eqnarray}
\label{dl}
\Delta_L&=& -\frac{\alpha}{4\pi}\ln \frac{M_p^2}{m_W^2} 
             \frac{V_{tb}^*V_{cb}m_b^2}{2m_W^2s_W^2}(1+\tan^2\beta)
             \left ( \tilde M^2_Q+\tilde M^2_D+\tilde M^2_{H_1}
                     +\vert A_d\vert^2 \right ),\\
\label{dr}
\Delta_R&=& -\frac{\alpha}{4\pi}\ln \frac{M_p^2}{m_W^2} 
             \frac{V_{tb}^*V_{cb}m_b^2}{2m_W^2s_W^2}(1+\tan^2\beta)\ 
             m_t A_d^*.
\end{eqnarray}
$\theta_t$ is the mixing angle\footnote{Note that our definition of 
$\theta_t$ differs from that in Ref.~\cite{hikasa} by a minus sign. }
between left- and right-handed stops,
defined by  
\begin{eqnarray}
\left( \begin{array}{l} \tilde t_1 \\   \tilde t_2 \end{array} \right)
=\left( \begin{array}{ll} \cos\theta_t & \sin\theta_t \\
           -\sin\theta_t & \cos\theta_t  \end{array} \right)
\left( \begin{array}{l} \tilde t_L \\   \tilde t_R \end{array} \right).
\end{eqnarray}

In the above, we have adopted the notation in Ref.~\cite{gunion}, with
$m_{\tilde t_1}<m_{\tilde t_2}$.
$\tilde M^2_Q$, $\tilde M^2_D$ and $\tilde M^2_{H_1}$ are soft-breaking
mass terms for left-handed squark doublet $\tilde Q$, right-handed down 
squark $\tilde D$ and Higgs doublet $H_1$, respectively. 
$A_d$ is the coefficient of the trilinear term $H_1 \tilde Q \tilde D$
in soft-breaking terms and $\tan\beta=v_2/v_1$ is ratio of the vacuum
expectation values of the two Higgs doublets. Note that $\tilde c_L$ is
a mass eigenstate in our analysis since we do not consider the mixing
between  $\tilde c_L$ and  $\tilde c_R$, which is proportional to
the charm quark mass.   

From the above equations, we note that besides the large logarithmic 
factor $\ln (M_p^2/m_W^2)$, the mixings are proportional to $\tan^2\beta$ 
and thus can be further enhanced at large $\tan\beta$.
If we assume that all soft SUSY-breaking parameters are of
the same orders in magnitude, we then have typically
$\epsilon\approx 0.01(\tan\beta/10)^2$ and thus $\epsilon$ 
is much smaller than unity. (Note that to make the approximate 
expansion of Eq.~(\ref{mix}) valid, $\epsilon$ should be much 
smaller than unity.) 
Without such an assumption, $\epsilon$  can be larger 
because in the sum 
$\tilde M^2_Q+\tilde M^2_D+\tilde M^2_{H_1}+\vert A_d\vert^2$
only  $\tilde M_Q$ is related to stop and scharm masses while other 
parameters are independently free in the MSSM. 

The stop mass $m_{\tilde t_1}$ is particularly important for our study and will be 
retained as a free parameter in our numerical calculations. 
The lightness of the stop is quite well motivated in some SUSY models 
like SUGRA and is also preferred by electroweak baryogenesis \cite{Carena}.
On the other hand, the current lower bound on its mass is about 
135 GeV  \cite{FCCC}, albeit under some assumptions.
We will thus explore the mass range 
\begin{equation}
\label{stop-range}
150\ {\gev} < m_{\tilde t_1}  < 250\ {\gev}
\end{equation}
where the upper end is the kinematic limit for a 500 GeV linear collider.
So we assume an upper bound of about 250 GeV in our numerical analysis.
  
The flavor mixing between stop and scharm will induce the FCNC 
stop decay $\tilde t_1 \to c \tilde \chi^0_1$. 
Since the charge-current decay $\tilde t_1\to   b \tilde \chi^+_1$ 
can be the other important decay mode, 
the branching ratio of the FCNC decay is obtained by 
\begin{eqnarray}
\label{bf}
BF=\frac{\Gamma(\tilde t_1 \to c \tilde \chi^0_1)}
           {\Gamma(\tilde t_1 \to c \tilde \chi^0_1)
        +  \Gamma(\tilde t_1 \to b \tilde \chi^+_1)}
\end{eqnarray}
with 
\begin{eqnarray}\label{width1}
& &\Gamma(\tilde t_1 \to c \tilde \chi^0_j)
=\frac{\alpha}{2} \vert \epsilon\vert^2
  m_{\tilde t_1} 
  \left (1-\frac{m^2_{\tilde \chi^0_1}}{m^2_{\tilde t_1}}\right )^2 \
\left\vert e_c N'_{j1}+\frac{1}{s_W c_W}(\frac{1}{2}-e_c s_W^2) N'_{j2}
\right \vert^2 \, ,\\
\label{width2}
& & \Gamma(\tilde t_1 \to b \tilde \chi^+_j)
=\frac{\alpha}{4} m_{\tilde t_1} 
\left (1-\frac{m^2_{\tilde \chi^+_1}}{m^2_{\tilde t_1}}\right )^2 \ 
\left |-\frac{V_{j1}^*}{s_W}\cos\theta_t 
 +\frac{m_t V_{j2}^*}{\sqrt{2} m_W s_W \sin\beta}\sin\theta_t\right |^2\, .
\end{eqnarray}
Here, $N'_{ij}$ denotes the matrix element projecting the $i$-th neutralino 
into photino ($j=1$), zino ($j=2$), and two neutral Higgsinos ($j=3,4$).     
$V_{ij}$ is the matrix element projecting the $i$-th left-handed chargino
into wino ($j=1$) and the charged Higgsino ($j=2$).
The gaugino masses and mixing are determined by the soft 
SUSY-breaking parameters $M_1, M_2$, as well as $\mu,\ \tan\beta$.
There are strong theoretical motivations to further constrain these 
parameters \cite{run2}.
First of all, the supergravity models predict the unification relation
$M_1=\frac{5}{3}M_2\tan^2\theta_W\simeq 0.5 M_2$. Radiative electroweak
symmetry breaking generally yields a large $\mu$ parameter, although the
naturalness arguments prefer a lower value of $\mu$. This
scenario leads to the LSP $\tilde \chi^0_1$ bino-like, and 
$\tilde \chi^+_1$ wino-like, which is also favored for a SUSY
dark matter interpretation. Regarding the other parameter $\tan\beta$,
the LEP experiments excluded small values $\tan\beta<2$ \cite{LEP_higgs}.
For the sake of illustraton, we thus choose the following representative 
set of parameters 
\begin{eqnarray} \label{para}
M_2=150 {\rm ~GeV},\  \mu=300 {\rm ~GeV},\  \tan\beta=10. 
\end{eqnarray}
The chargino and neutralino masses in units of GeV are then given by
\begin{eqnarray}
& & m_{\tilde\chi^+_1}=133,\  ~m_{\tilde\chi^+_2}=328,~\nonumber\\
\label{char-mass}
& & m_{\tilde\chi^0_1}=72,\ ~m_{\tilde\chi^0_2}=134,\ 
~m_{\tilde\chi^0_3}=308,\  ~m_{\tilde\chi^0_4}=327.
\end{eqnarray}
In our analysis the chargino
$\tilde \chi^+_1$ must be lighter than stop $\tilde t_1$. Such a light
chargino decays into $f f' \tilde \chi^0_1$ ($f$ is a quark or lepton)
through exchanging a $W$-boson, or a charged Higgs boson, 
a slepton, a squark \cite{hikasa2}. 
Since, typically, the charged Higgs, sleptons and squarks are much heavier 
than the $W$-boson, such decays occur dominantly through the 
$W$-exchange diagram and the branching ratio for the clean
channels $\tilde \chi^+_1\to \ell^+ \nu\chi^0$ ($\ell=e$ and $\mu$) 
is thus approximately 2/9. 

With the parameters in Eq.~(\ref{para}), the branching fraction 
$B(\tilde t_1 \to c \tilde \chi^0_1)$ in the no mixing limit 
is approximately given by
\begin{eqnarray}
BF \approx \left \{ \begin{array} {ll}
  1.3   \vert \epsilon\vert^2, &{~~{\rm for}~ m_{\tilde t_1}=150~\gev } , \\
  0.16 \vert \epsilon\vert^2,& { ~~{\rm for}~ m_{\tilde t_1}=250~\gev }.
\end{array} \right.  
\end{eqnarray}
For a lighter $m_{\tilde t_1}$, the decay  $\tilde t_1 \to b \tilde \chi^+_1$
is kinematically suppressed; and for a heavier $m_{\tilde t_1}$, 
this charged-current channel becomes dominant. 

Note that our
choice of parameters in Eq.~(\ref{para}) is rather representative
for which the decay modes $\tilde t_1\to c\tilde \chi^0_1$
and $b \tilde \chi^+_1$ are both kinematically accessible.
The exception is in the Higgsino-like region ($M_2>|\mu|$).
In this case, both the LSP and $\tilde \chi^+_1$ are mainly 
Higgsino-like, and are about degenerate in mass close to $\mu$.
The lepton produced in the decay 
$\tilde \chi^+_1\to \tilde\chi^0_1 \ell^+ \nu$ will be too soft to 
be experimentally identifiable, making the signal difficult to
observe. As we indicated earlier, this situation is disfavored
by the arguments of SUSY-GUT and dark matter. We will thus not
pursue this special case further.

\section{ Observability of  FCNC Stop Decay at Colliders}

Since the stop $\tilde t_1$ is likely to be significantly
lighter than any other squark and thus the production rate
of $\tilde t_1 \tilde{\bar t_1}$ is larger than other squark 
pairs, as well as than $\tilde t_1 \tilde{\bar t_2}$ or 
$\tilde t_2 \tilde{\bar t_2}$,  we only consider the 
production of $\tilde t_1 \tilde{\bar t_1}$ in our analysis.
Inclusion of the channels $\tilde t_1 \tilde{\bar t_2}$ and 
$\tilde t_2 \tilde{\bar t_2}$ would enhance the signal observability
although the kinematics of the final states would be more involved to study.
For a light stop with a mass close to the top
quark, the QCD corrections enhance the total cross section of stop pair 
by a factor of about $1.2$ at the Tevatron energy and $1.4$ at the LHC 
energy~\cite{zerwas}. This enhancement (the so-called $K$ factor)
will be taken into account 
in our calculation. The one-loop corrections to stop pair production in 
a 500 GeV $e^+e^-$ collider were found to increase the cross section by 
10--20\%\ \cite{stop_NLC} and we assume an enhancement 
factor $K=1.1$ in our analyses. 
Going beyond the crude assumption on the branching fractions of the
$\tilde t_1$ decay in the Tevatron studies \cite{FCNC,FCCC}, we consider
the FCNC decay of one stop $\tilde{\bar t_1} \to \bar c 
\tilde \chi^0_1$, and the charge-current decay of the other one,
$\tilde t_1\to b \tilde \chi^+_1  \to b \ell^+ \nu \tilde\chi^0_1$.
The signal we are proposing to look for is a  
$\tilde t_1 \tilde{\bar t_1}$ event giving rise to
an energetic isolated charged lepton ($e$ or $\mu$), 
a $b$-quark jet, a (charm) jet and missing transverse energy,
denoted by $j b \ell+\etmiss$. 

First, we consider the search at hadron colliders. To simulate the acceptance 
of the detectors, we impose some
kinematical cuts on the transverse momentum ($p_{T}$), the pseudo-rapidity
($\eta$), and the separation in the azimuthal angle-pseudo rapidity plane 
($\Delta R= \sqrt{(\Delta \phi)^2 +(\Delta \eta)^2})$ between 
a jet and a lepton or between two jets. We choose the basic
acceptance cuts for the Tevatron
\begin{eqnarray}
\nonumber
&&p_T^{\ell},\ p_T^{\rm jet},\ {\etmiss}\ge 20 {\rm~GeV},\\ 
&&\eta_{\rm jet},\ \eta_{\ell} \le 2.5,\\
&&\Delta R_{jj},~\Delta R_{j\ell} \ge 0.5.
\nonumber
\end{eqnarray}
We increase the threshold for the LHC as
\begin{eqnarray}
\nonumber
&&p_T^{\ell}\ge 20~ {\gev},\ p_T^{\rm jet}\ge 35~ {\gev},\ 
{\etmiss}\ge 30 {\rm~GeV},\\ 
&&\eta_{\rm jet},\ \eta_{\ell} \le 3,\\
&&\Delta R_{jj},~\Delta R_{j\ell} \ge 0.4.
\nonumber
\end{eqnarray}
Furthermore, we assume the  tagging of a $b$-quark jet
with $50 \%$ efficiency and the probability of 0.4\% (15\%) 
for a light quark ($c$-quark) jet to be mis-identified as a $b$-jet.

To make the analyses more realistic, we simulate the energy resolution
of the calorimeters by assuming a Gaussian smearing on the energy 
of the final state particles, given by
\begin{eqnarray}
&&{\Delta E \over E} = 
{30\% \over \sqrt{E}} \oplus 1\% \quad {\rm for\ leptons},\\
&&{\Delta E \over E} = 
{80\% \over \sqrt{E}} \oplus 5\% \quad {\rm for\ hadrons},
\end{eqnarray}
where $E$ is in GeV.

The potential SM backgrounds at hadron colliders are
\begin{itemize}
\item[1)] $bq (\bar q) \rightarrow tq'(\bar q')$; 
\item[2)] $q\bar q' \rightarrow W^* \rightarrow t\bar b$;
\item[3)] $Wb\bar b;\  Wc\bar c;\ Wcj;\ Wjj$;
\item[4)] $t\bar t\rightarrow W^-W^+b\bar b$;
\item[5)] $gb\rightarrow tW$;
\item[6)] $qg\rightarrow q't\bar b$.
\end{itemize}
The backgrounds 5) and 6) are of modest rates and can mimic our
signal only if the extra jet is missed in detection. After vetoing the
extra central jet, these backgrounds are effectively suppressed. 
The $t\bar t$ background 4) is of a large production cross
section, especially at the LHC. It can mimic our signal if 
both $W$'s decay leptonically and one of the charged leptons is not 
detected, which is assumed to occur if
the lepton pseudo-rapidity and transverse momentum are
in the range $\vert\eta(\ell)\vert>3$ and $p_T(\ell)<10$ GeV.
In addition, we also have some SUSY backgrounds. In case of $\tilde t_1$
being significantly lighter than other squarks, the dominant SUSY background
is the pure charged-current decay
$\tilde t_1 \tilde{\bar t_1} \to \tilde \chi^+_1 \tilde \chi^-_1 b \bar b$.
Also, at the high end of the top squark mass range considered in our analysis,
$\tilde t_1$ can decay to $t^* + \tilde \chi^0_1$ or even $t + \tilde \chi^0_1$.
All these processes give a $t\bar t$-like signature \cite{stop_mc} and
 can mimic our signal just like the SM production of $t\bar t$.
However, compared with $t\bar t$ background 4), 
such backgrounds are much smaller since their production rates are much lower
than the $t\bar t$ background.

\begin{table}[tb]
\begin{center}
\begin{tabular}{|c|c|c|c|c|}
        &\multicolumn{2}{c|}{Tevatron 2 TeV} &\multicolumn{2}{c|}{LHC 14 TeV} \\ \cline{2-5}
        & ~basic~cuts &~basic~+~$m_T$ & ~basic cuts~ & ~basic~+~$m_T$\\\hline

signal                       &  23 & 6.6  & 720  & 220  \\\hline
~~$qb\to q^\prime t$~~       &  120  & 5.0 & 7400 & 400  \\\hline
$q\bar q^\prime \to t\bar b$ &  39 & 2.3 & 280  & 17   \\\hline
$Wb\bar b$                   &  130  & 2.5 & 570  & 45   \\\hline
$Wc\bar c$                   &  80   & 1.5 & 450  & 36   \\\hline
$Wcj$                        &  670  & 3.8 & 7600 & 650  \\\hline
$Wjj$                        &  500  & 2.9 & 1700 & 150  \\\hline
$t\bar t$                    &  7.9  & 3.8 & 600  & 300  \\
\end{tabular}
\vskip 0.3cm
\caption{Signal  $\tilde t_1 \bar{\tilde t_1}\to \ell b c\  \etmiss$ 
and background cross sections in units of fb. 
The signal results were calculated by assuming $m_{\tilde t_1}=150$ GeV
and other parameters are in Eq.~(\ref{para}).
The charge conjugate channels have been included. 
The signal results do not include the branching fraction factor $2 BF(1-BF)$,
which should be multiplied to obtain the actual signal rate for a given value 
of $BF$.
\label{tab1} }
\end{center}
\end{table}

We notice that for most of the background events the missing energy comes 
only from neutrinos in $W$ decay, while for the signal the missing energy 
contains the extra contribution from the neutralinos.
From the transverse momentum of the lepton ${\vec p}_T^{\ell}$ 
and the missing transverse momentum $\vec \ptmiss$,
we construct the transverse mass as  
\begin{equation}
m_T = \sqrt{ (|\vec p_T^{\ell}|+|\vec \ptmiss|)^2
- (\vec p_T^{\ell}+\vec \ptmiss)^2}.
\end{equation}
For the background events where the only missing energy comes from 
a neutrino from $W$ decay, $m_T$ is always less than $M_W$ (and peaks 
just below $M_W$) without energy smearing. Smearing pushes some of the
events above $M_W$. 
For the signal $m_T$ is spread out widely above and below $M_W$, 
due to the extra missing energy of the neutralinos. 
In order to substantially enhance the signal-to-background ratio ($S/B$), 
we apply a cut
\begin{equation}
m_T > 90 {\rm~ GeV}.
\end{equation}

\begin{figure}[tb] 
\centerline{\psfig{file=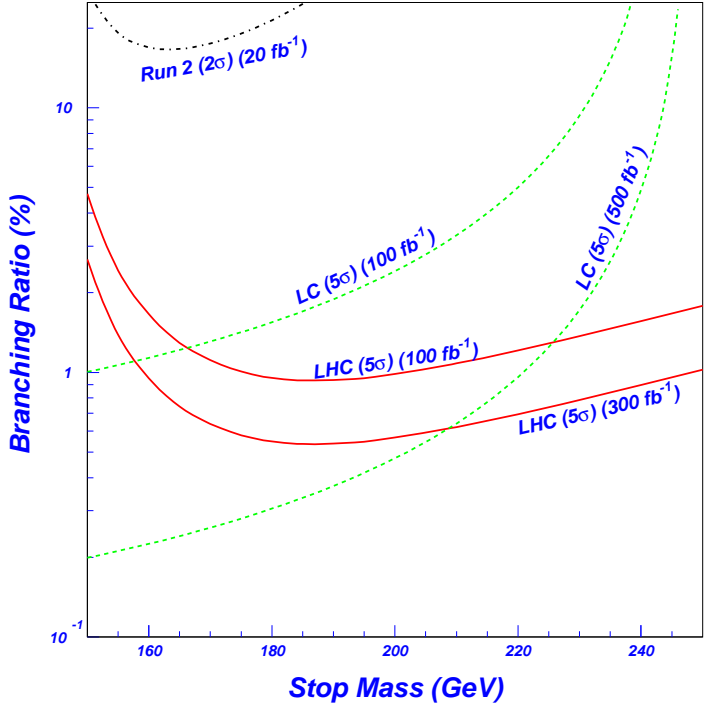,width=4in,angle=0} }
\vspace*{0.5cm}
\caption{Discovery limits of the branching ratio of the FCNC stop decay
 $\tilde t_1 \to c \tilde \chi^0_1$  versus the stop mass.  The region above 
each curve is the corresponding observable region. }
\label{fig1} 
\end{figure}

We first present the signal and background cross sections at the
Tevatron (2 TeV) and LHC (14 TeV) under various cuts in Table \ref{tab1}. 
One sees that with only the basic acceptance cuts, the various SM 
backgrounds can overwhelm the signal. The implementation of 
the $m_T$ cut reduces the  backgrounds $Wb\bar b$, 
$Wc\bar c$, $Wjj$ and  $Wcj$  efficiently.  
The production rate of the signal 
$b$+jet$+\ell\ \etmiss$ can be obtained by multiplying the 
 $\tilde t_1 \tilde{\bar t_1}$ cross section 
by the branching fraction factor $({2}/{9})2BF(1-BF)$ for
a given value of $BF$, 
Our results for the $2\sigma$  sensitivity on  $BF$
at the Tevatron versus the stop mass are shown in Fig.~\ref{fig1} (the top curve).
We find that due to limited statistics, Tevatron
Run 2 with a luminosity of 2 fb$^{-1}$ is not able to discover 
the signal nor even set any significant  bounds on the branching 
fraction $BF$.  A bound at $2\sigma$ level could be reached at the Tevatron energy
with a luminosity of 20 fb$^{-1}$ for $m_{\tilde t_1}<180$ GeV,
corresponding to $BF\sim 20\%$.
At the LHC the $5\sigma$-discovery is accessible, reaching the branching fraction
below $1\%$ even for a low luminosity 100 fb$^{-1}$.
From Fig.~\ref{fig1} one sees that the detection sensitivity for hadron colliders 
does not monotonously increase as the stop mass
decreases. Instead, when the stop becomes too light, the detection sensitivity
decreases. This is the effect of the cuts applied in our
simulation and can be understood as follows. As the stop mass decreases, 
the stop pair production rate increases. However, when the stop becomes too light, 
the $b$-jet from $\tilde t_1\to \tilde \chi^+_1 b$ becomes very soft 
and thus failed to pass the selection cuts so that it decreases the 
detection sensitivity. 

The results at the LHC are obtained by applying the basic and the $m_T$ 
cuts.  The signal significance is  obtained in terms of Gaussian statistics,
given by the signal and background events $S/\sqrt B$.
Although the sensitivity reach at the LHC is impressive as shown in Fig.~\ref{fig1},
the signal-to-background ratio $S/B$ becomes rather low when reaching the
small branching fraction. Thus the sensitivity relies on the successful control of
the systematics in the experiments.

\begin{figure}
\centerline{\psfig{file=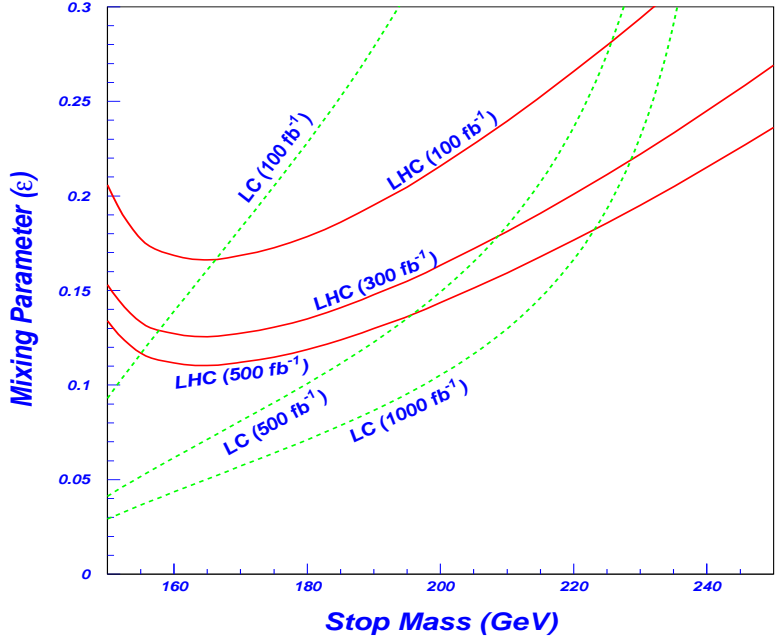,width=11cm,height=9cm,angle=0} }
\vspace*{0.5cm} 
\caption{ $5\sigma$ discovery limits of the stop-scharm mixing parameter $\epsilon$
versus the stop mass with the stop mixing angle $\theta_t=\pi/10$. 
 The region above each curve  is the corresponding observable region.}
\label{fig2}
\end{figure}

It is known that the experimental environment is much  cleaner at an 
$e^+e^-$ collider. 
Now we recapitulate our analyses for an $e^+e^-$ linear collider with 
C.~M.~energy of 500 GeV.
Since the environment of $e^+e^-$ colliders is much cleaner, 
we will evaluate the production rate of the signal 
$b$+jet$+\ell\ \etmiss$ simply by multiplying the cross section 
$\sigma(e^+e^-\to \tilde t_1 \tilde{\bar t_1})$, 
the branching ratio $({4}/{9})BF(1-BF)$, 
the $b$-tagging efficiency assumed to be $50\%$, 
and the detection efficiency of kinematics assumed to be $80\%$.
The possible SM backgrounds are 
\begin{eqnarray}
& & e^+e^-\to W^+W^-\to j j' \ell \nu,\\
& & e^+e^-\to t\bar t\to bW^+ \bar b W^-\to  b\bar b jj' \ell \nu.
\end{eqnarray} 
However, these backgrounds can be effectively separated due to
the rather different kinematical features from the signal.
If we define the recoil mass as
\begin{equation}
m_r^2 = (P_{e^+}+P_{e^-}- \sum P_{\rm obs})^2,
\end{equation}
where the sum is over all momenta of the observed final state
particles, then we notice that the backgrounds have rather
small recoil mass from the single missing neutrino.
The recoil mass for the signal on the other hand
is quite large since the neutralino is very massive. 
Similar to the case of hadron colliders, the dominant SUSY background
is $\tilde t_1 \tilde{\bar t_1} 
\to \tilde \chi^+_1 \tilde \chi^-_1 b \bar b$, which is small
and neglected in our numerical analysis. For a stop $m_{\tilde t_1}\lsim 230$ GeV
when the threshold is sufficiently open for $\sqrt s=500$ GeV, one can reach
a $5\sigma$ observation at the LC with a branching fraction of 1--5\%\ 
even for a  luminosity of only 100 fb$^{-1}$, as shown in Fig.~\ref{fig1}.

\begin{figure}
\centerline{\psfig{file=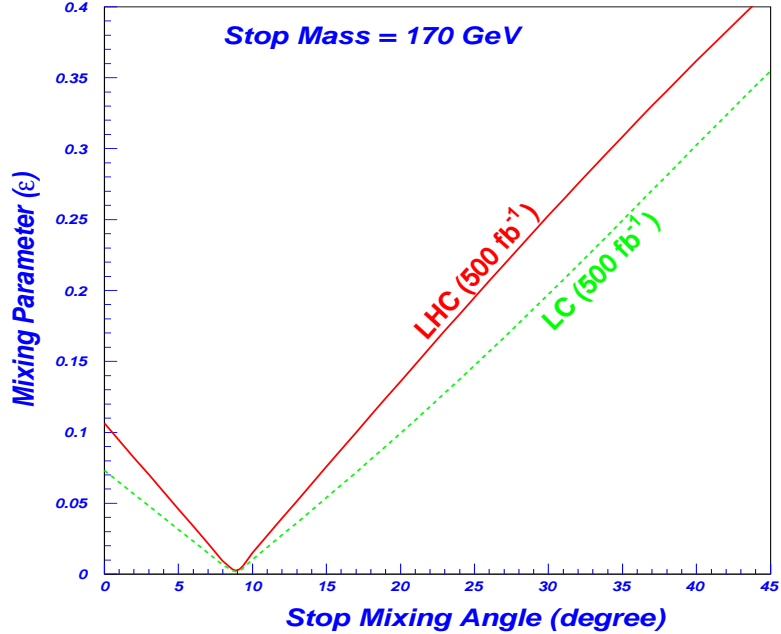,width=11cm,height=9cm,angle=0} }
\vspace*{0.5cm} 
\caption{ $5\sigma$ discovery limits of the stop-scharm mixing parameter $\epsilon$
versus the stop mixing angle $\theta_t$ for $m_{\tilde t_1}= 170$ GeV.
The region above each curve  is the corresponding observable region.}
\label{fig3}
\end{figure}

The exclusion and discovery limits of the branching fraction can be translated
into the limits on the stop-scharm mixing parameter $\epsilon$, which can
be predicted for a specific SUSY model. At this stage, the 
stop mixing angle $\theta_t$ needs
to be specified.  For illustration, we first fix 
$\theta_t=\pi/10$ and the resulting limits on $\epsilon$ are shown in Fig.~\ref{fig2}, 
corresponding to the results of Fig.~\ref{fig1}.  For stop mass of 150 GeV
the $5\sigma$ discovery limit with a luminosity of 100 fb$^{-1}$
is $\epsilon \gsim 0.09$ at the LC, and 
$\epsilon \gsim 0.20$ at the LHC. For a heavier stop, the 
detection sensitivity at the LC drops much more rapidly than that at the
LHC due to the limited C.M. energy of the LC.  
Note that although Run 2 with 20 fb$^{-1}$ has a $2\sigma$ sensitivity 
(as shown in Fig.~\ref{fig1}) to the decay branching ratio, it has no sensitivity 
to  $\epsilon<0.3$. When the  $2\sigma$ sensitivity limits of Run 2 in
Fig.~\ref{fig1} are translated to the mixing parameter $\epsilon$, rather large 
$\epsilon$ values ($\gsim 0.5$) are obtained.  
A large $\epsilon$ is not theoretically favored, as implied in Eqs.~($1-4$).

The obtained limits on $\epsilon$ are sensitive to the mixing angle $\theta_t$,
which controls the partial width for the CC decay as seen in Eq.~(\ref{width2}).
The dependence of $\epsilon$ limits on $\theta_t$ is 
shown in Fig.~\ref{fig3} for a fixed value of stop mass  $m_{\tilde t_1}= 170$ GeV.
For the mixing angle to be near a certain value, 
$\tan\theta_t \approx (\sqrt2 m_W \sin\beta/m_t) (V_{11}/V_{12}) $, 
the CC mode is suppressed and the sensitivity to the FCNC mode is greatly
enhanced. For our choice of the SUSY parameters, this occurs near 
$\theta_t\approx 9^\circ$.  The $5\sigma$ sensitivity with 500 fb$^{-1}$ for
the LHC and LC could reach as far as $\epsilon\approx 0.01$. For nearly
maximal mixing $\theta_t\approx 45^\circ$, on the other hand, the sensitivity
could be reduced to about $\epsilon\approx 0.4$.  

Furthermore, the limits or observation of the mixing parameter $\epsilon$ can be 
translated into some knowledge  on certain soft SUSY breaking parameters. 
From Eqs.~(\ref{eps}) and (\ref{dl}), we see that the mixings are proportional
to a sum of certain  parameters, typically like  
$(\tilde M_Q^2+\tilde M_D^2+\tilde M_{H_1}^2+|A_d|^2)/(m_{\tilde t}^2-m_{\tilde c_L}^2)$.
This can vary independently  of 
$m_{\tilde t}$  and  $m_{\tilde c_L}$ since only $\tilde M_Q$ 
in this sum is related to  $m_{\tilde t}$  and  $m_{\tilde c_L}$.
For the purpose of illustration,  taking $m_{\tilde t_1}=150$ GeV, $\theta_t=\pi/10$, 
$m_{\tilde c_L}=200$ GeV and  other SUSY parameters given in Eq.~(\ref{para}), 
we obtain the LHC discovery ($5\sigma$) limit 
with a luminosity of 100 fb$^{-1}$
\begin{eqnarray}
\label{soft-limit1}
\sqrt{\tilde M_Q^2+\tilde M_D^2+\tilde M_{H_1}^2+|A_d|^2+0.3 m_t A_d^*  } \gsim 1.4 {\rm~TeV},
\end{eqnarray} 
or, in case of non-observation, the $2\sigma$ bound given by 
\begin{eqnarray}
\label{soft-limit2}
\sqrt{\tilde M_Q^2+\tilde M_D^2+\tilde M_{H_1}^2+|A_d|^2+0.3 m_t A_d^* }
\lsim 1.0 {\rm~TeV}.
\end{eqnarray} 
Due to the nature of the multiple parameters as a combination 
involved in the expression, more comprehensive
analyses would be needed, possibly to combine with other experimental knowledge
on the SUSY parameters, in order to extract the information for the theory parameter
space.

\section{Conclusions} 

In summary, we studied the potential of detecting 
the FCNC stop decay $\tilde t_1 \to c \tilde \chi^0_1$, as a probe of
stop-scharm mixing, at the upgraded Tevatron, the LHC and the LC. 
Rather than performing an exhaustive scan of the  SUSY parameters, 
we chose a representative set of the relevant parameters to
demonstrate the possibility of observation.  
Through Monte Carlo simulation,
we found that the signal at the Tevatron is too weak to be observable
for the choice of well-motivated SUSY parameters.
At the LHC on the other hand,  with judicial kinematical cuts, 
it is quite possible to observe a $5\sigma$
signal with a branching fraction as low as $1\%$ even for a luminosity of 
100 fb$^{-1}$.  However, it should be noted that systematic effects in the
experiments must be under control.
At  an LC of $\sqrt s=500$ GeV, one can reach a $5\sigma$ observation 
with a branching fraction of $1-5\%$ for  a luminosity of 100 fb$^{-1}$.
The limits or observation of this important decay mode can be 
translated into some knowledge  on certain soft SUSY breaking parameters. 
We finally note that in our study we have chosen a representative scenario for relevant 
SUSY parameters in which the lightest neutralino and chargino
are gaugino-like. In the region of SUSY parameters where 
the lightest neutralino and chargino are Higgsino-like,
the signal would  be more difficult to observe.

\bigskip
\noindent{\bf Acknowledgments}

The work of T.H.  is supported in part by the US Department of Energy 
under grant DE-FG02-95ER40896,
in part by the Wisconsin Alumni Research Foundation, and in part by
National Natural Science Foundation of China (NNSFC). The work of K.H. is 
supported in 
part by the Grant-in-Aid for Scientific Research (No.~12640248 and 14046201) 
from the Japan Ministry of Education, Culture, Sports, Science, and 
Technology.  X.Z. is supported by NNSFC.


\begin{thebibliography}{99}

\bibitem{fcp} A. Masiero and O. Vives, Ann.~Rev.~Nucl.~Part.~Sci.~{\bf 51}, 161 (2001).
\bibitem{hpz} R.~D.~Peccei and X.~Zhang, \NPB337, 269 (1990);
              T.~Han, R.~D.~Peccei and X.~Zhang, \NPB454, 527 (1995).
\bibitem{susyf} For a comprehensive analysis, see, {\it e.~g.},
                S. Dimopoulos and D. Sutter, \NPB452, 496 (1996);
                F. Gabbiani, E. Gabrielli, A. Masiero, and L. Silvestrini,
                \NPB477, 321 (1996); and references therein.
\bibitem{tcvh_sm} For top FCNC decays in the SM, see,
            G.~Eilam, J.~L.~Hewett and A.~Soni, \PRD44, 1473 (1991);
            B.~Mele, S.~Petrarca, A.~Soddu, \PLB435, 401 (1998).
\bibitem{tcv_mssm}  For $t \to cV$ in the MSSM, see,  {\it e.~g.},
                 C.~S.~Li, R.~J.~Oakes and J.~M.~Yang, \PRD49, 293 (1994); 
                 G.~Couture, C.~Hamzaoui and H.~Konig, \PRD52, 1713 (1995);
                 J.~L.~Lopez, D.~V.~Nanopoulos and R.~Rangarajan, \PRD56, 3100  (1997);
                 G.~M.~de Divitiis, R.~Petronzio and L.~Silvestrini, \NPB504, 45 (1997);
                 J.~M.~Yang, B.-L.~Young and X.~Zhang, \PRD58, 055001 (1998);
                 J. Cao, Z. Xiong and J. M. Yang, \NPB651, 87 (2003).
\bibitem{tch_mssm}  For $t \to ch$ in the MSSM, see,  {\it e.~g.},
                    J.~M.~Yang and C.~S.~Li, \PRD49, 3412 (1994); 
                    J.~Guasch and J.~S\`{o}la, \NPB562, 3 (1999);
                    S.~B\'{e}jar, J.~Guasch and J.~S\`{o}la, hep-ph/0101294;
                    G.~Eilam, {\it et al.}, \PLB510, 227 (2001); 
                    J.~L.~Diaz-Cruz, H.-J.~He, and C.-P.~Yuan,  Phys.~Lett.~{\bf B530}, 179 (2002).
\bibitem{duncan} See, {\it e.~g.}, M. J. Duncan, \NPB221, 285 (1983).
\bibitem{hall} See, {\it e.~g.}, N.~Arkani-Hamed, H.-C.~Cheng, J.~L.~Feng 
               and L.~J.~Hall, \PRL77, 1937 (1996); \NPB505, 3 (1997);
               J.~Cao, T.~Han, X.~Zhang and G.~Lu, \PRD59, 095001 (1999).
\bibitem{review} For a review, see, {\it e.~g.}, 
                 M.~Misiak, S.~Pokorski, J.~Rosiek, hep-ph/9703442.
\bibitem{Endo} M.~Endo, M.~Kakizaki, and M.~Yamaguchi, hep-ph/0311072. 
\bibitem{hikasa} K. Hikasa and M. Kobayashi, \PRD36, 724 (1987).
\bibitem{FCNC} 
   D0 Collaboration:  S.~Abachi {\it et al.},  Phys.~Rev.~Lett.~{\bf 76}, 2222 (1996);
   CDF Collaboration: T.~Affolder {\it et al.}, Phys.~Rev.~Lett.~{\bf 84}, 5704 (2000).
\bibitem{stop-3-body} W. Porod, T. Wohrmann, \PRD55, 2907 (1997); 
                      W. Porod, \PRD 59, 095009  (1999);
                      A. Djouadi and  Y. Mambrini, \PRD63, 115005  (2001).
\bibitem{FCCC} 
   D0 Collaboration: V.~M.~Abazov {\it et al.}, Phys.~Rev.~Lett. {\bf 88}, 171802 (2002);
   CDF Collaboration: D.~Acosta {\it et al.}, Phys.~Rev.~Lett. {\bf 90}, 251801 (2003).
\bibitem{hosch}   S. Mrenna and C.-P. Yuan, \PLB367, 188 (1996);
                  M. Hosch {\it et al.,} \PRD58, 034002 (1998).
\bibitem{cdf-tt} CDF Collaboration: T.~Affolder {\it et al.}, \PRD63, 091101 (2001).
\bibitem{gunion}  H. E. Haber and G. L. Kane, Phys. Rep. {\bf 117}, 75 (1985);
                  J. F. Gunion and H. E. Haber, \NPB272, 1 (1986).
\bibitem{Carena}
	M.~Carena, M.~Quiros and C.E.~Wagner,
	\PLB380, 81 (1996); \NPB503, 387 (1997);
	\NPB524,  3 (1998);
	D.~Delepine, J.M.~Gerard, R.~Gonzalez Felipe and J.~Weyers,
	\PLB386, 183 (1996); J.~McDonald, \PLB413, 30 (1997);
	J.M.~Cline and G.D.~Moore, \PRL181, 3315 (1998).
\bibitem{run2} See, {\it e.~g.}, V.~Barger, C.E.M.~Wagner {\it et al.},
Tevatron Run-II Workshop report, hep-ph/0003154.
%
\bibitem{LEP_higgs} See, {\it e.~g.}, 
                     P.~A.~McNamara and S.~L.~Wu, Rept.~Prog.~Phys.~{\bf 65}, 465 (2002).
\bibitem{hikasa2} K. Hikasa and T. Nagano, \PLB435, 67 (1998);
                  N. Oshimo and Y. Kizukuri, \PLB186, 217 (1987).
\bibitem{zerwas}  W.~Beenakker, M.~Kramer, T.~Plehn, M.~Spira, and 
                  P.~M.~Zerwas, \NPB515, 3 (1998).
\bibitem{stop_NLC}  H. Eberl, A. Bartl, and W. Majerotto, \NPB472, 481 (1996);
                    X.-J. Bi, Y.-B. Dai, and X.-Y. Qi, \PRD62, 115004 (2000).
\bibitem{stop_mc}  R. Demina, J. D. Lykken, K. T. Matchev and A. Nomerotski, 
\PRD62, 035011 (2000); 
                   J.~M.~Yang and B.-L.~Young, \PRD62, 115002 (2000);
                   A. Djouadi, M. Guchait, Y. Mambrini, \PRD64, 095014 (2001); 
                   E. L. Berger and T. M. P. Tait, hep-ph/0002305.
\end{thebibliography}
\end{document}